\documentclass[aps,prd,twocolumn,groupedaddress,amssymb,eqsecnum,showpacs,epsfig,nofootinbib]{revtex4}
\usepackage{graphicx}
\usepackage{bm}
\usepackage{dcolumn}
\usepackage{amsmath}

\usepackage{amsmath}
\usepackage{amssymb}

\numberwithin{equation}{section}
\usepackage{epsfig}

\def \lleq {\lower0.9ex\hbox{ $\buildrel < \over \sim$} ~}
\def \ggeq {\lower0.9ex\hbox{ $\buildrel > \over \sim$} ~}

\def \omm  {\Omega_{0 {\rm m}}}

\def \beq  {\begin{equation}}
\def \eeq  {\end{equation}}
\def \ber  {\begin{eqnarray}}
\def \eer  {\end{eqnarray}}

\begin{document}
\newcommand{\newc}{\newcommand}

\newc{\be}{\begin{equation}}
\newc{\ee}{\end{equation}}
\newc{\ba}{\begin{eqnarray}}
\newc{\ea}{\end{eqnarray}}
\newc{\bea}{\begin{eqnarray*}}
\newc{\eea}{\end{eqnarray*}}
\newc{\D}{\partial}
\newc{\ie}{{\it i.e.} }
\newc{\eg}{{\it e.g.} }
\newc{\etc}{{\it etc.} }
\newc{\etal}{{\it et al.}}
\newc{\lcdm }{$\Lambda$CDM }
\newcommand{\nn}{\nonumber}
\newc{\ra}{\rightarrow}
\newc{\lra}{\leftrightarrow}
\newc{\lsim}{\buildrel{<}\over{\sim}}
\newc{\gsim}{\buildrel{>}\over{\sim}}
\title{Bright High z SnIa: A Challenge for LCDM?}
\author{L. Perivolaropoulos$^a$ and A. Shafieloo$^{b,c}$}
 \affiliation{$^a$Department of Physics, University of Ioannina, Greece \\$^b$Department of Physics, University of Oxford, 1 Keble Road, Oxford, OX1 3NP, UK\\$^c$BIPAC, University of Oxford, Denys Wilkinson Building, 1 Keble Road, Oxford OX1 3RH, UK}
\date{\today}

\begin{abstract}
It has recently been pointed out by Kowalski et. al. (arxiv:0804.4142) that there is `an unexpected brightness of the SnIa data at $z>1$'. We quantify this statement by constructing a new statistic which is applicable directly on the Type Ia Supernova (SnIa) distance moduli. This statistic is designed to pick up systematic brightness trends of SnIa datapoints with respect to a best fit cosmological model at high redshifts. It is based on binning the normalized differences between the SnIa distance moduli and the corresponding best fit values in the context of a specific cosmological model (eg $\Lambda CDM$). These differences are normalized by the standard errors of the observed distance moduli. We then focus on the highest redshift bin and extend its size towards lower redshifts until the Binned Normalized Difference (BND) changes sign (crosses 0) at a redshift $z_c$ (bin size $N_c$). The bin size $N_c$ of this crossing (the statistical variable) is then compared with the corresponding crossing bin size $N_{mc}$ for Monte Carlo data realizations based on the best fit model. We find that the crossing bin size $N_c$ obtained from the Union08 and Gold06 data with respect to the best fit \lcdm model is anomalously large compared to $N_{mc}$ of the corresponding Monte Carlo datasets obtained from the best fit \lcdm in each case. In particular, only $2.2\%$ of the Monte Carlo \lcdm datasets are consistent with the Gold06 value of $N_c$ while the corresponding probability for the Union08 value of $N_c$ is $5.3\%$. Thus, according to this statistic, the probability that the high redshift brightness bias of the Union08 and Gold06 datasets is realized in the context of a $(w_0,w_1)=(-1,0)$ model (\lcdm cosmology) is less than $6\%$. The corresponding realization probability in the context of a $(w_0,w_1)=(-1.4,2)$ model is more than $30\%$ for both the Union08 and the Gold06 datasets indicating a much better consistency for this model with respect to the BND statistic.
\end{abstract}
\pacs{98.80.Es,98.65.Dx,98.62.Sb}
\maketitle

\section{Introduction}
The discovery of the accelerating expansion of the universe about a decade ago \cite{SN} has led to an intensive pursue of the physical origin of this acceleration. This pursue has been taking place in both the observational and the theoretical aspects of the problem.

On the theoretical aspect, there has been significant progress made by pointing out several models that may produce the observed accelerating expansion and clarifying the limits of their predictions with respect to the observed expansion rate as a function of redshift. For example it has been pointed out that theoretical models based on modifications of general relativity \cite{theory-cross}, interacting dark energy \cite{Das:2005yj} or higher dimensional brane world models \cite{brane} can easily predict an effective dark energy equation of state $w(z)$ that crosses the Phantom Divide Line (PDL) $w=-1$. On the other hand, models based on general relativity that are free from instabilities\cite{pdl-inst} and conserve energy and momentum of dark energy have a $w(z)$ confined in the range $w(z)\geq -1$.

On the observational aspect there has been significant improvement of the constraints on the recent Hubble expansion history $H(z)$ coming from a diverse set of cosmological observations. Such observations include direct geometrical probes (standard candles like SnIa \cite{SN,sn2}, gamma ray bursts \cite{GRB} and standard rulers like the CMB sound horizon\cite{Percival:2007yw,Komatsu:2008hk}) and dynamical probes (growth rate of cosmological perturbations \cite{growth} probed by the redshift distortion factor or by weak lensing \cite{weak-lens}).

All these observational probes are converging towards confirming the accelerating expansion of the universe assuming the homogeneity of the universe. They have ruled out at several $\sigma$ a flat matter dominated universe (assuming a power-law form of the primordial spectrum) and they have produced excellent fits for the simplest cosmological model predicting accelerating cosmic expansion. This model is based on the presence of the cosmological constant $\Lambda$ and Cold Dark Matter (\lcdm)\cite{lcdm-rev}.

In view of the significant present and forecasted improvement of relevant cosmological observations there are specific theoretical questions that are becoming particularly interesting. For example the question {\it `Is general relativity the correct theory on cosmological scales?'} is particularly interesting but perhaps premature for the current status of observational data which still allow a considerable range of $w(z)$ forms around the simplest allowed value $w=-1$ corresponding to \lcdm. A more realistic but equally important question for the current status of observational data is the following: {\it `Is \lcdm the correct model of the accelerating expansion of the universe?'} This `yes-no' question is more realistic because \lcdm is a well defined model which makes clear and definite predictions that are easily falsifiable. On the other hand violations of general relativity can often be mimicked by (peculiar) properties of dark energy such as anisotropic stress or clustering\cite{Amendola:2007rr}.

Most approaches in answering the above question for the validity of \lcdm have focused on comparing \lcdm with alternative models or parameterizations on the basis of a bayesian analysis. Due to its simplicity and acceptable quality of $\chi^2$ fit, \lcdm usually comes out as a winner in such a comparison \cite{bayes-lcdm} even though certain potential problems of the model on small \cite{Peebles:2003pk} and large scales \cite{probs-lcdm} have recently been identified.


An alternative approach discussed in the present study is to directly compare the real data with Monte Carlo simulations consisting of fictitious cosmological data that would have been obtained in the context of a \lcdm cosmology. This comparison can be made on the basis of various statistics which attempt to pick up features of the data that can be reproduced with difficulty by a \lcdm cosmology.

The existence of such features is hinted by the form of the likelihood contours in various parameter planes containing parameter values corresponding to flat $\Lambda$CDM. For example, most SnIa datasets producing likelihood contours in the $\Omega_\Lambda - \omm$ parameter plane have the $1\sigma$ contour barely intersect the line of flatness  $\Omega_\Lambda + \omm = 1$ at the lower left side of the contour \cite{gold06,union08}\footnote{Even though the first year SNLS data did not have this feature, there are preliminary indications that this feature will appear in the three year SNLS data\cite{rppc}}. Similarly, likelihood contours based on either SnIa standard candles or standard rulers (CMB sound horizon or Baryon Acoustic Oscillations) and constraining the parametrization \cite{Chevallier:2000qy} \be w(z)=w_0+ w_1 \frac{z}{1+z} \label{cpl} \ee systematically have the point corresponding to \lcdm $(w_0,w_1)=(-1,0)$ at the lower right edge of the $1\sigma$ contour while the best fit involves $w_0 <-1$, $w_1 >0$ \cite{gold06,union08,cross-evid,data-cross}. This feature has persisted consistently over the last decade and over different accelerating expansion probes \cite{cross-evid} (SnIa standard candles and CMB-BAO standard rulers). Even though the statistical significance of these features when viewed individually is relatively low, their persistent appearance makes it likely that there are systematic differences between the cosmological data and \lcdm predictions.

One such difference in the context of SnIa data has been recently pointed out by Kowalsky et. al. \cite{union08} where it was stated that there is `an unexpected brightness of SnIa data at $z>1$'. This feature is even directly visible by observing the SnIa distance moduli superposed with the best fit \lcdm model (dashed line in Fig. 1) where most high $z$ moduli are below the best fit \lcdm curve (obviously the reverse happens at low redshifts to achieve a good fit). Notice that this bias is smaller in the context of a parametrization that crosses the PDL $w=-1$ (continuous line in Fig. 1). In the PDL crossing model we fix $w_0$, $w_1$ and vary $\omm$ only, in order to mimic the \lcdm number of parameters.

\begin{figure}[!t]
\hspace{0pt}\rotatebox{0}{\resizebox{.5\textwidth}{!}{\includegraphics{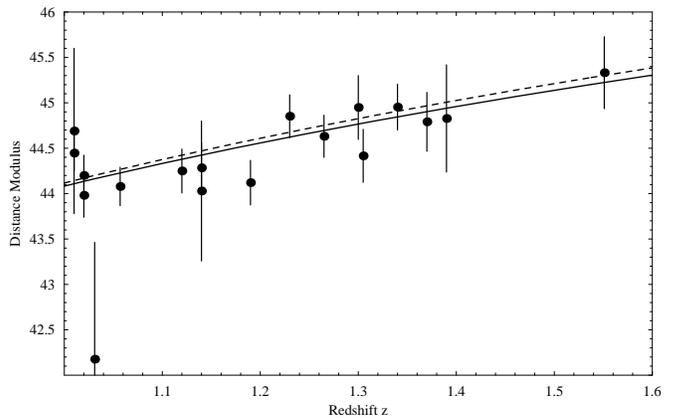}}}
\vspace{0pt}{\caption{The Union08\cite{union08} distance moduli data superposed with the best fit \lcdm model ($\omm=0.29$) dashed line and with the best fit  $(w_0,w_1)=(-1.4,2)$ model ($\omm=0.30$) continous line. Notice that at high redshifts z the distance moduli tend to be below the \lcdm  best fit while the trend is milder in the PDL crossing best fit model.}} \label{fig1}
\end{figure}

 This anomalous behavior of the data with respect to the \lcdm best fit may be attributed to the systematic brightness trend of high redshift SnIa with respect to the best fit \lcdm model. It is likely that this bias of the SnIa data with respect to \lcdm best fit is also responsible for the systematic mild preference (at $1\sigma$) of the SnIa data for a $w(z)$ crossing the $w=-1$ line. The goal of the present paper is to study quantitatively the likelihood of the existence of the above described bias in the context of a \lcdm cosmology. For this reason we use a statistic (the Binned Normalized Differences (BND)) specially designed to pick up systematic brightness trends of the SnIa data with respect to a best fit cosmological model at high redshift.

\begin{figure*}[ht]
\centering
\begin{center}
$\begin{array}{@{\hspace{-0.10in}}c@{\hspace{0.0in}}c}
\multicolumn{1}{l}{\mbox{}} &
\multicolumn{1}{l}{\mbox{}} \\ [-0.2in]
\epsfxsize=3.3in
\epsffile{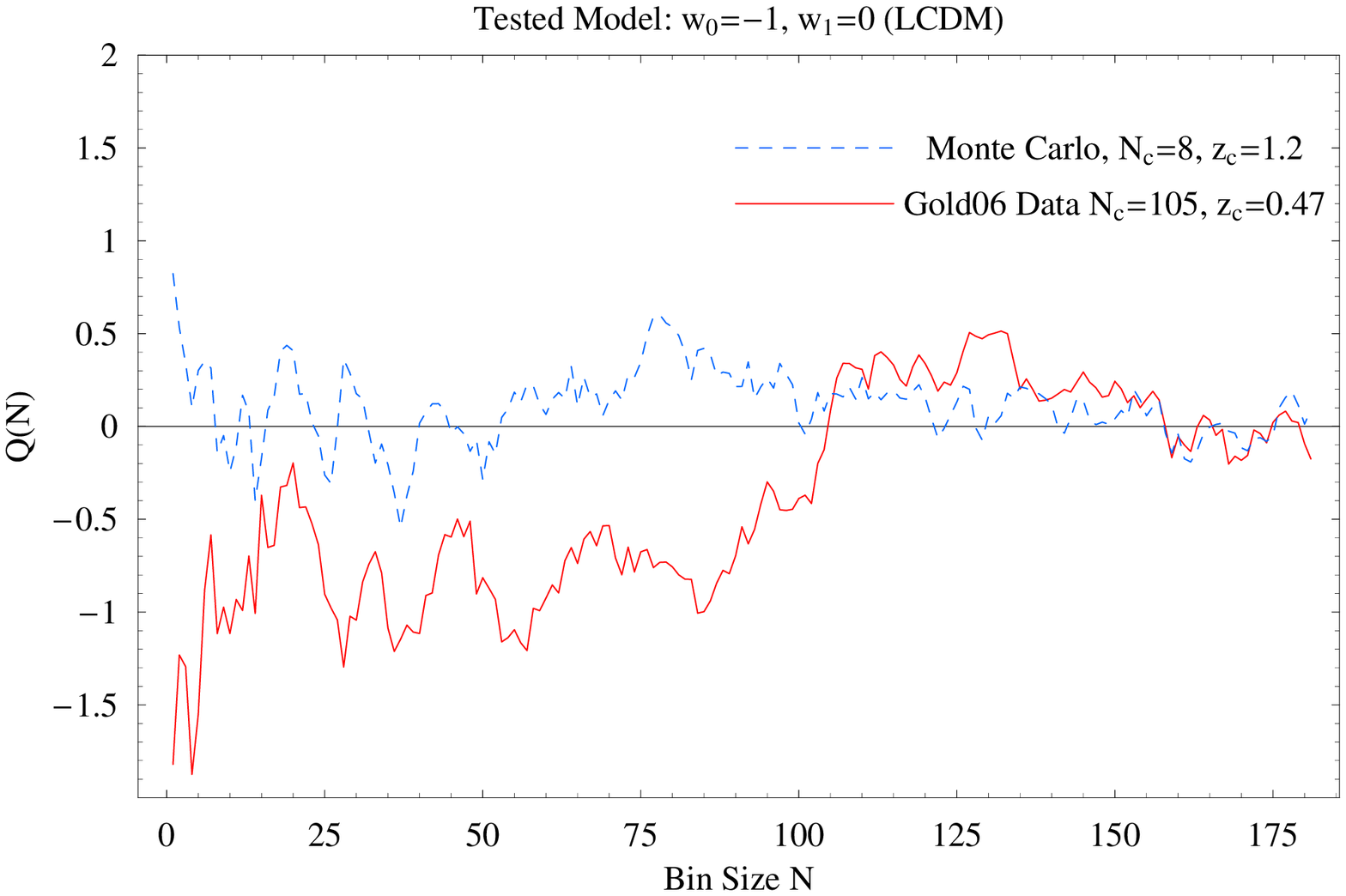} &
\epsfxsize=3.3in
\epsffile{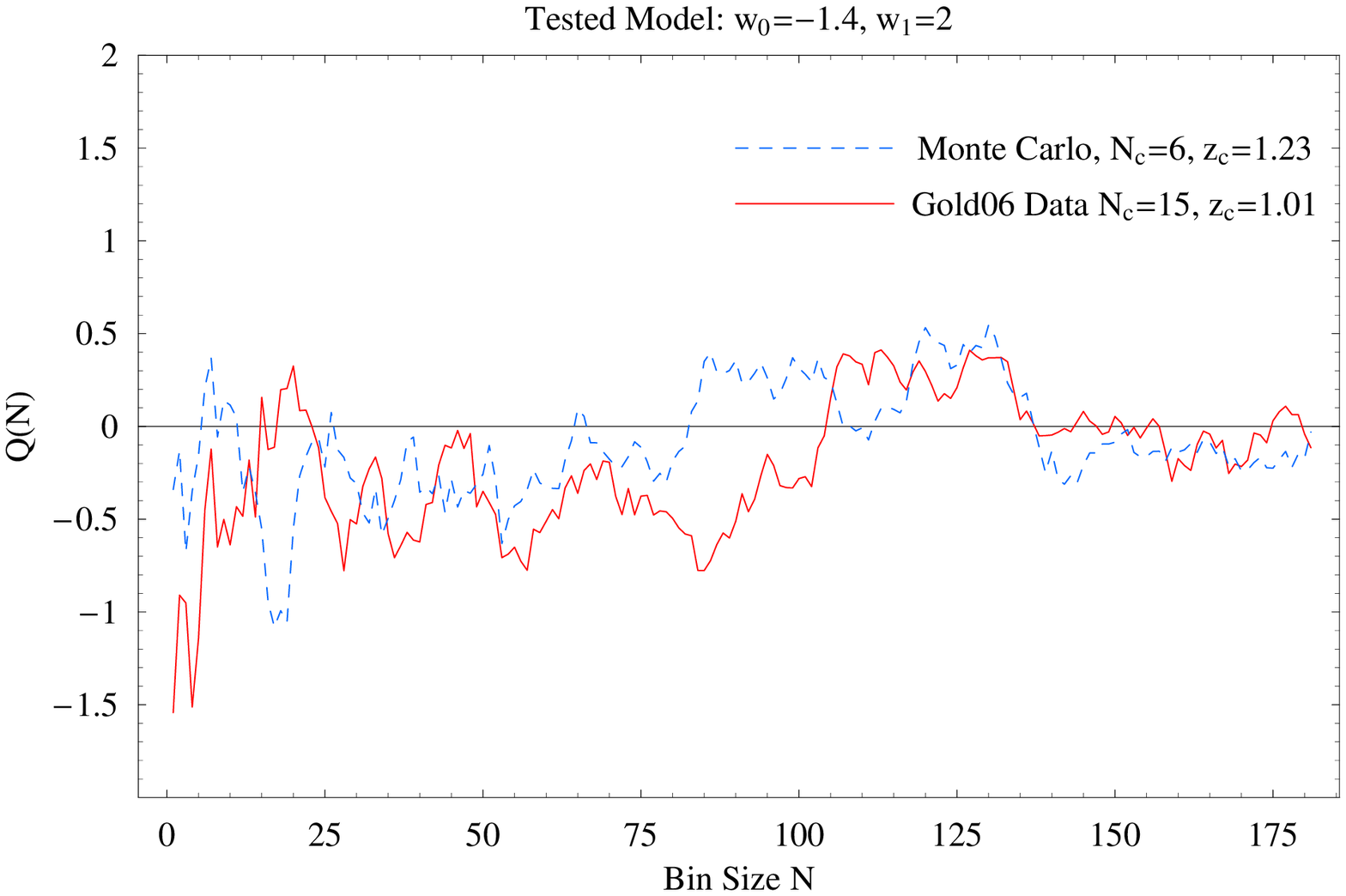} \\
\end{array}$
\end{center}
\vspace{0.0cm}
\caption{\small a: The form of the Binned Normalized Difference $Q(N)$ obtained from the Gold06\cite{gold06} dataset assuming \lcdm at
best fit ($\omm=0.34$) (red line) along with a typical form of $Q(N)$ obtained from Monte Carlo data based on the best fit \lcdm and the Gold06
dataset (blue dashed line). b: Similar plot testing the PDL crossing model $(w_0,w_1)=(-1.4,2)$ (best fit $\omm=0.34$) instead of \lcdm. The agreement between real and Monte Carlo data is significantly better in this case.}
\label{fig2}
\end{figure*}

\section{The Binned Normalized Difference (BND) Statistic}

The main advantages of the BND statistic discussed in this section may be summarized as follows:
\begin{enumerate}\item It is directly applicable on the distance moduli data. \item It is a `yes-no' statistic for each model and it involves no comparison with alternative models or parametrizations. Such comparisons introduce new uncertainties involving the suitability of the chosen parametrization in the infinite dimensional functional space of possible parameterizations. For example a parametrization choice of $w(z)=w_0=constant$ leads to $w_0 = -1\pm 0.1$ \cite{union08} favoring \lcdm more than the CPL choice of eq. (\ref{cpl}) which leads to $w_0=-1.1\pm 0.3$, $w_1=1\pm 2$ \cite{union08}. \item It focuses on specific features of the data with respect to the best fit model (systematic brightness trends) thus exposing clearly the weak points of the model. \item It is insensitive to the uncertainties of the matter density in the sense that it does not require fixing $\omm$ to a value motivated from other cosmological observations or even marginalizing with respect to it to smooth out the dependence on $\omm$.  In order to test LCDM (or the other similar one parameter parametrizations) we just  minimize $\chi^2$ with respect to $\omm$, find the best fit $d_L(z)$ and then apply the BND statistic. No external input or marginalization on $\omm$ is needed.
In contrast, in the maximum likelihood method with multiparameter parametrizations, the choice of the matter density, $\omm$ can significantly affect the reconstructed properties of dark energy \cite{matter}.\end{enumerate}
\begin{figure*}[ht]
\centering
\begin{center}
$\begin{array}{@{\hspace{-0.1in}}c@{\hspace{0.0in}}c}
\multicolumn{1}{l}{\mbox{}} &
\multicolumn{1}{l}{\mbox{}} \\ [-0.20in]
\epsfxsize=3.3in
\epsffile{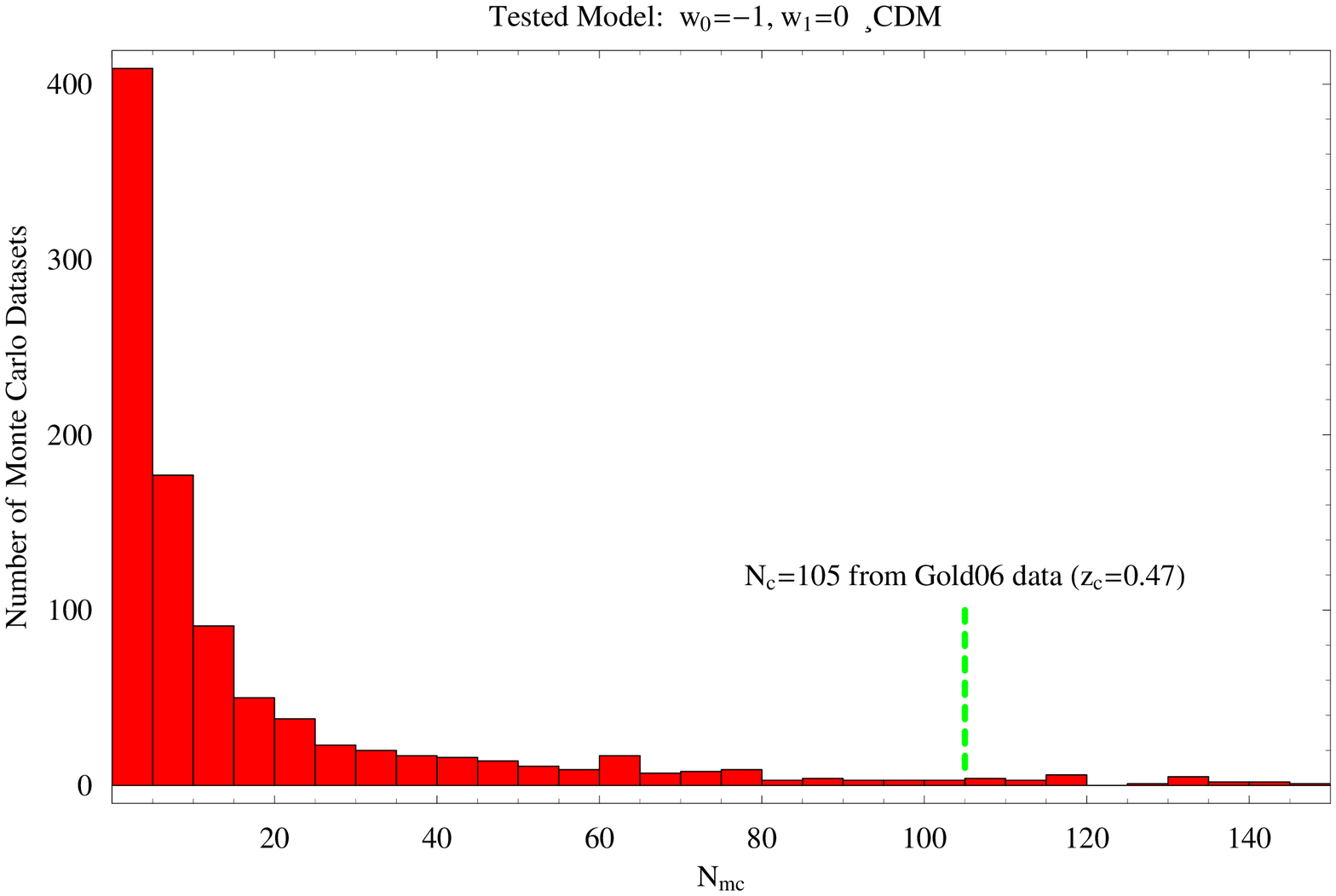} &
\epsfxsize=3.3in
\epsffile{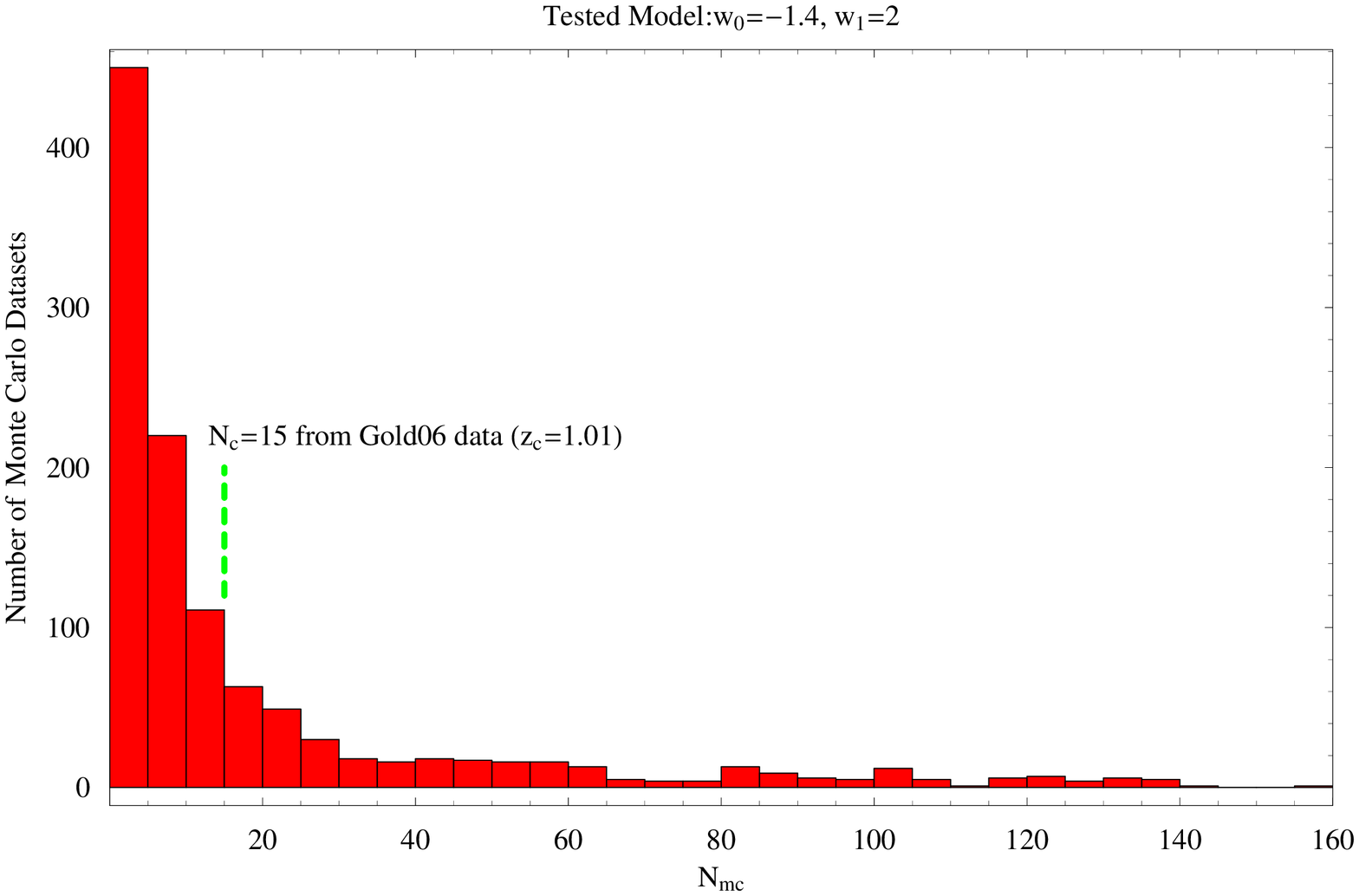} \\
\end{array}$
\end{center}
\vspace{0.0cm}
\caption{\small a: A histogram of the probability distribution of $N_{mc}$ obtained using Monte Carlo \lcdm data ($\omm=0.34$) in the context of the Gold06\cite{gold06} dataset. The thick green dashed line
corresponds to the crossing redshift $z_c$ of the real Gold06 data. b: Similar histogram for the PDL crossing model $(w_0,w_1)=(-1.4,2)$
(best fit $\omm=0.34$) instead of \lcdm. Notice that the crossing redshift $z_c$ corresponding to the real Gold06 data is a much more probable event in
the context of this cosmological model.}
\label{fig3}
\end{figure*}
\begin{figure*}[ht]
\centering
\begin{center}
$\begin{array}{@{\hspace{-0.1in}}c@{\hspace{0.0in}}c}
\multicolumn{1}{l}{\mbox{}} &
\multicolumn{1}{l}{\mbox{}} \\ [-0.20in]
\epsfxsize=3.3in
\epsffile{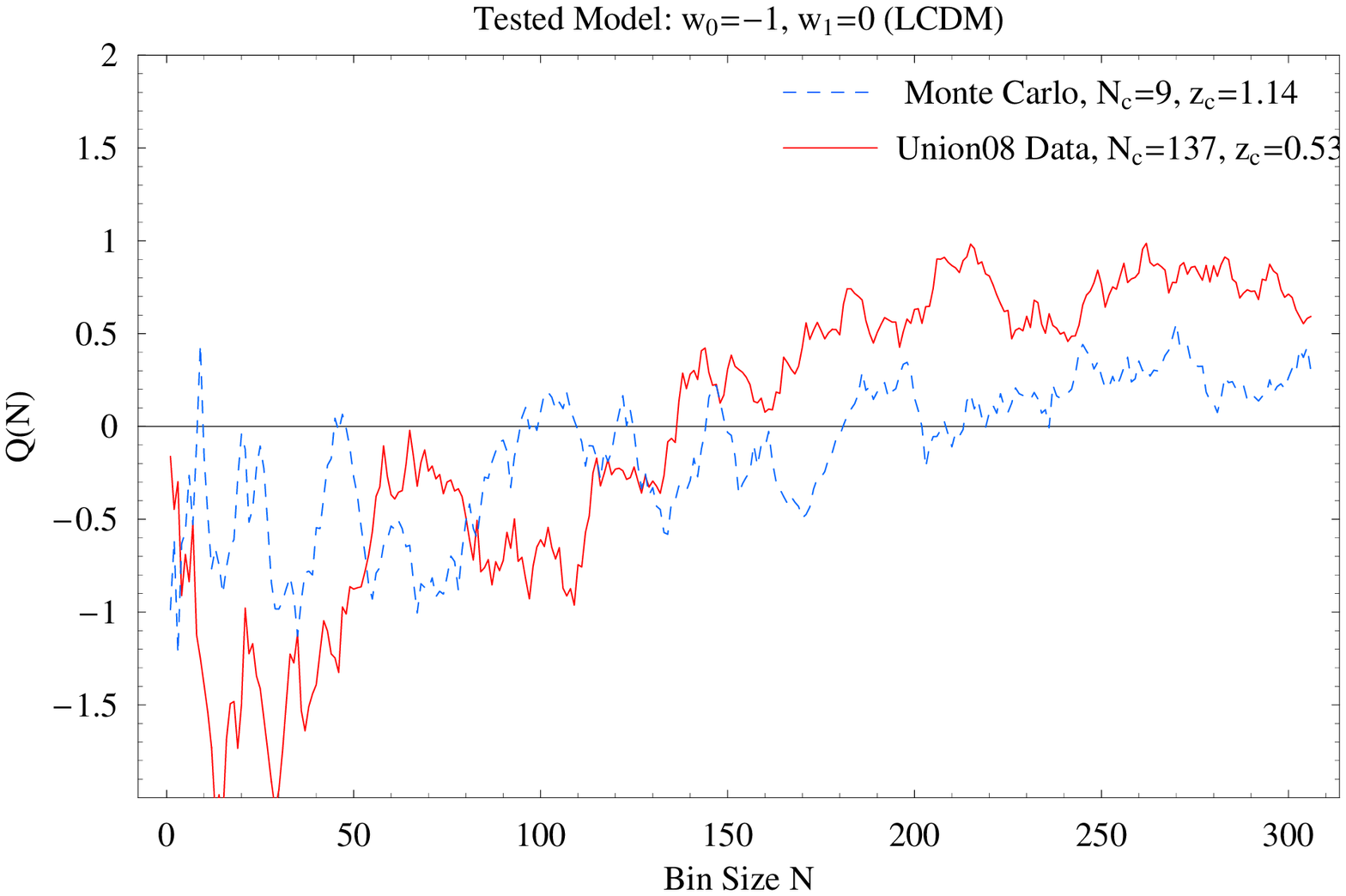} &
\epsfxsize=3.3in
\epsffile{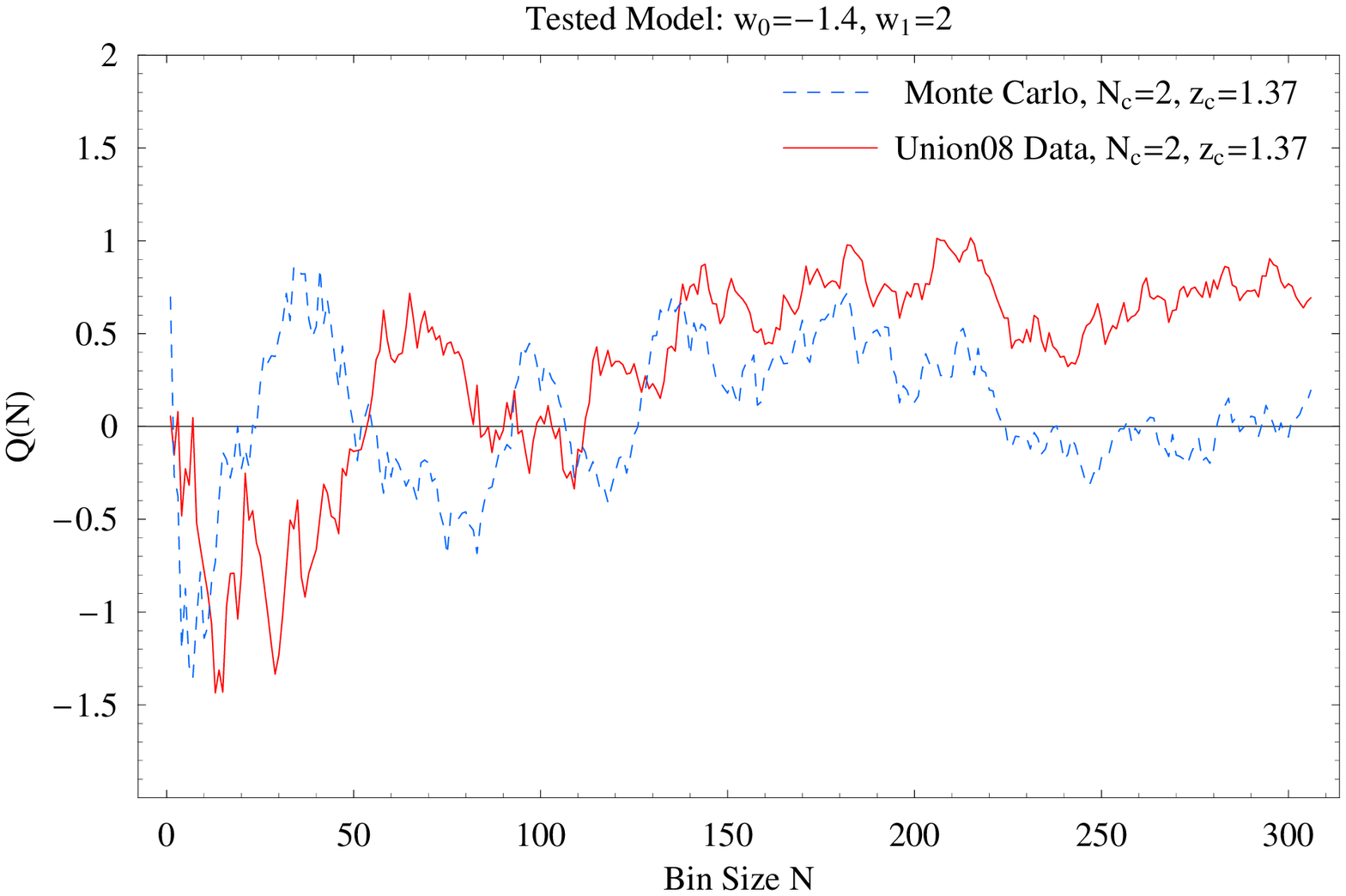} \\
\end{array}$
\end{center}
\vspace{0.0cm}
\caption{\small a: The form of the Binned Normalized Difference $Q(N)$ obtained from the Union08 dataset assuming \lcdm at best fit ($\omm=0.29$) (red line) along with a typical form of $Q(N)$ obtained from Monte Carlo data based on the best fit \lcdm and the Union08
dataset (blue dashed line). b: Similar plot testing the PDL crossing model $(w_0,w_1)=(-1.4,2)$ (best fit $\omm=0.30$) instead of \lcdm.}
\label{fig4}
\end{figure*}
The construction of our statistical analysis involves the following steps:
\begin{enumerate}
\item
Consider an $H(z)$ parametrization (eg \lcdm) to be tested and a SnIa distance moduli dataset $\mu_i(z_i)$ (eg Gold06\cite{gold06} or Union08\cite{union08}).\item
Obtain the best fit form ${\bar H (z)}$ of $H(z)$ and the corresponding distance moduli best fit $\bar \mu(z)$. \item
Construct the error normalized difference of the data from the best fit $\bar \mu(z)$ as \be q_i(z_i)=\frac{\mu_i(z_i) - {\bar \mu(z_i)}}{\sigma_i (z_i)} \label{qidef}\ee \item
Consider the highest redshift bin of the normalized differences $q_i (z_i)$ consisting of $N$ points $q_i(z_i)$ defined as the Binned Normalized Difference (BND) \be Q(N)=C_{N}\sum_{i=1}^N q_i(z_i) \label{qndef} \ee where $C_N=\frac{1}{\sqrt{N}}$ is a normalization factor (of no particular interest for our purpose), $z_1$ is the highest redshift of the dataset and the redshifts decrease in sequence as $i$ increases ($z_N$ is the lowest redshift of the sum). Obviously $Q(N)=Q(N(z_N))$ ie the BND variable Q is also expressible in terms of the minimum redshift $z_N$ of the sum (\ref{qndef}). Notice that due to the central limit theorem, $Q(N)$ is to a good approximation a gaussian random variable even if the luminosity distance errors are non-gaussian. Notice that $Q(N)<0$ implies that the SnIa in the redshift range $z_1=z_{max}>z>z_N$ are on the average brighter than the \lcdm prediction. The statistical significance of this additional brightness however requires comparison with Monte Carlo data (see 6 below). \item
We increase the bin size $N$ until $Q(N)$ changes sign for the first time at $N=N_c$, $z=z_c$. We consider $N_c$ (or equivalently $z_c$) as our statistical variable. Notice that if there are systematic brightness trends at high redshifts then $N_c$ is expected to be anomalously large (or equivalently $z_c$ anomalously low). \item
Finally, we ask how often does the value $N_c$ (or $z_c$) occur in Monte Carlo SnIa datasets produced from the considered best fit cosmological $H(z)$ parametrization. In particular for each Monte Carlo dataset we find the best fit form of $H(z)$ and follow the above steps in order to find the corresponding BND crossing redshift $z_{mc}$. We then find the fraction of Monte Carlo datasets leading to $z_{mc} \leq z_c$ (or equivalently $N_{mc} \geq N_c$). This fraction is a representation of the probability that the dataset would be realized in the context of the particular cosmological model. The Monte Carlo realization of a given datapoint at redshift $z_i$ in the context of a particular cosmological model, is a random gaussian variable with mean value $\bar \mu (z_i)$ (the best fit distance modulus at the given redshift) and standard deviation $\sigma_i(z_i)$ (the standard error of the corresponding datapoint).
\end{enumerate}
Even though we have found that the BND statistic is particularly efficient in picking up systematic brightness trends at high redshifts it clearly does not consist a unique approach. It is possible to construct other more complicated statistics aiming at testing brightness trends or other features of the distance moduli data. For our purpose however which is to quantify the high redshift brightness trend of the data, the BND statistic is sufficient since it combines effectiveness with simplicity.

\begin{figure*}[ht]
\centering
\begin{center}
$\begin{array}{@{\hspace{-0.10in}}c@{\hspace{0.0in}}c}
\multicolumn{1}{l}{\mbox{}} &
\multicolumn{1}{l}{\mbox{}} \\ [-0.20in]
\epsfxsize=3.3in
\epsffile{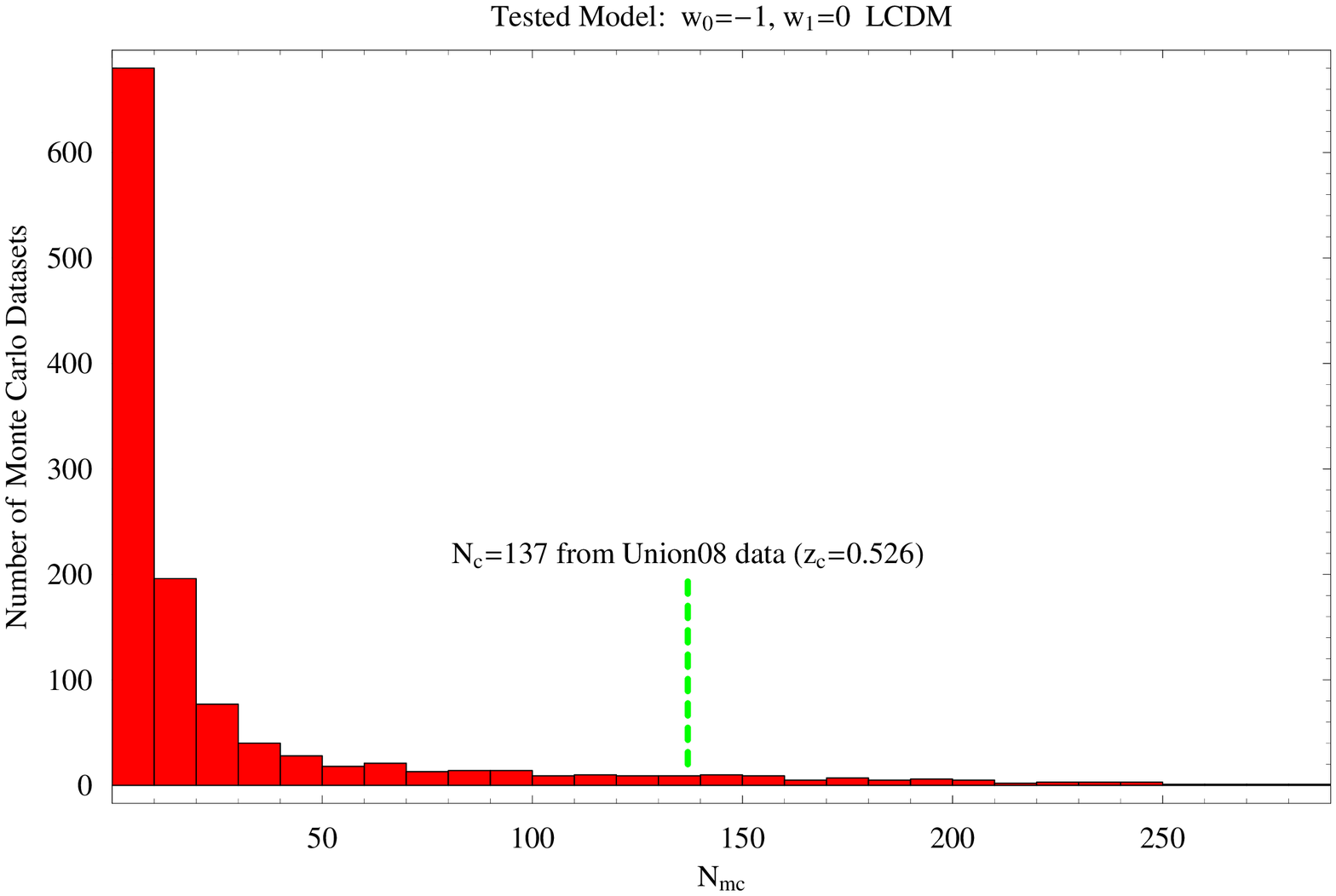} &
\epsfxsize=3.3in
\epsffile{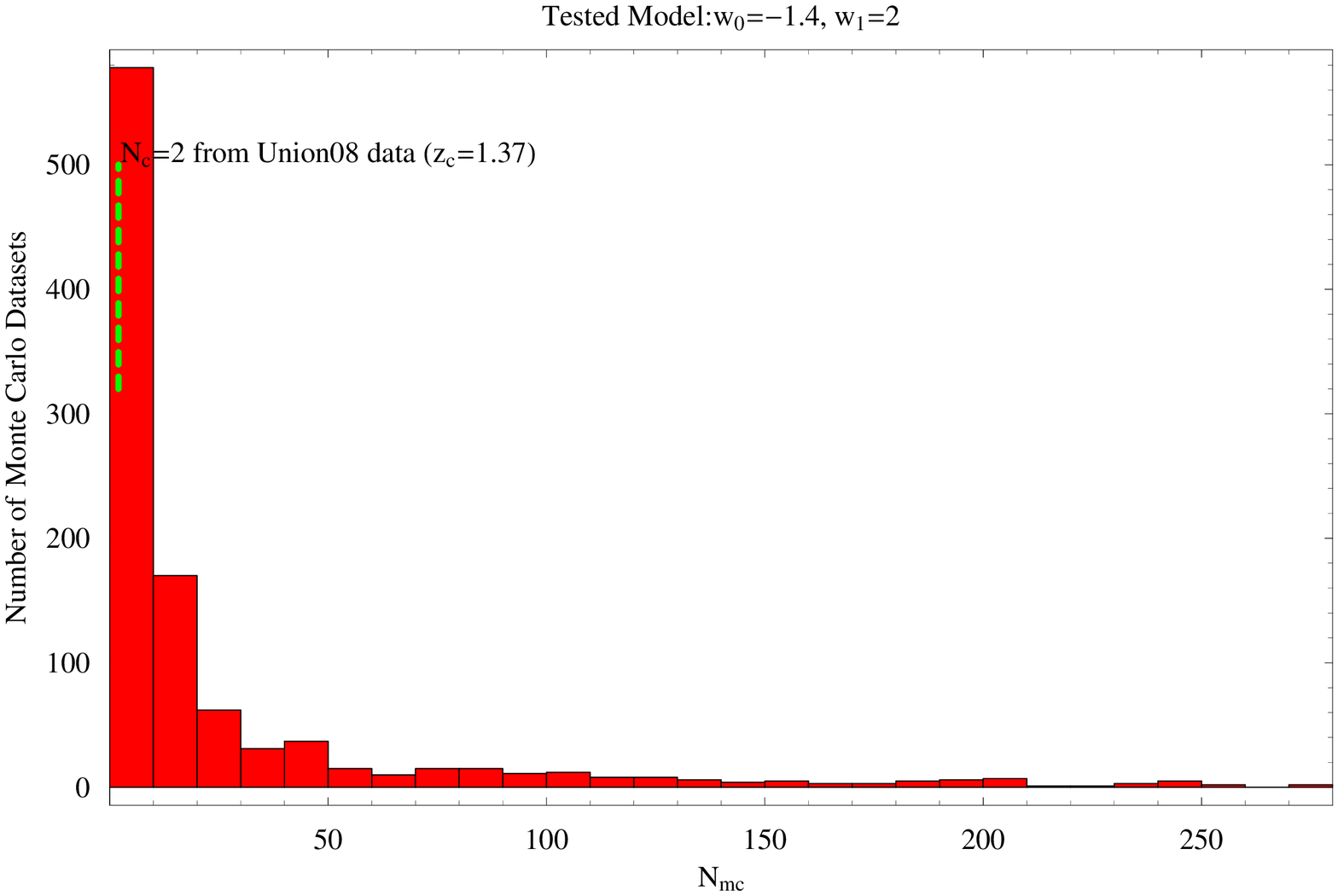} \\
\end{array}$
\end{center}
\vspace{0.0cm}
\caption{\small a: A histogram of the probability distribution of $N_{mc}$ obtained using Monte Carlo Union08 data under the assumption of a best fit \lcdm cosmological model ($\omm=0.29$). The thick green dashed line corresponds to the crossing redshift $z_c$ of the real Union08 data. Fig. b: Similar histogram for the PDL crossing model $(w_0,w_1)=(-1.4,2)$
(best fit $\omm=0.30$) instead of \lcdm. Notice that the crossing redshift $z_c$ corresponding to the real data is a much more probable event in  the context of this cosmological model.}
\label{fig5}
\end{figure*}

\section{Results}

We have applied the BND crossing statistic to both the Gold06\cite{gold06} and the Union08\cite{union08} datasets testing two cosmological models:
\begin{itemize}
\item \lcdm ($w_0=-1$, $w_1=0$ in eq. (\ref{cpl})).
\item A phantom divide crossing model ($w_0=-1.4$, $w_1=2$ in eq. (\ref{cpl})).
\end{itemize}
For each model and each dataset we first obtain the best fit value of $\omm$ and the best fit form of $\bar \mu(z)$ as in \cite{cross-evid}. We then construct $q_i(z_i)$ (eq. (\ref{qidef})) and $Q(N)$ (eq. (\ref{qndef})) and find $N_c$, $z_c$. Next we construct 1000 Monte Carlo datasets of distance moduli for each pair (model - dataset) and use them to obtain the corresponding values of $N_{mc}$, $z_{mc}$.

In Fig. 2a we show the form of $Q(N)$ obtained from the Gold06\cite{gold06} dataset (182 datapoints) assuming \lcdm at best fit ($\omm=0.34$) along with a typical form of $Q(N)$ obtained from Monte Carlo data based on the best fit \lcdm and the Gold06 dataset. Notice that the real data BND crossing redshift $z_c = 0.47$ ($N_c = 105$) is significantly lower (larger) than the corresponding redshift (bin size) obtained from the particular Monte Carlo dataset ($z_{mc}=1.2$, $N_{mc}=8$). The fraction of Monte Carlo datasets that can mimic a BND crossing redshift similar to the real data ($z_{mc} \leq z_c$, $N_{mc}\geq N_c$) is $2.2\%$. In Fig. 3a we show a histogram of the probability distribution of $z_{mc}$ obtained using Monte Carlo \lcdm data. The thick dashed line corresponds to the crossing redshift $z_c$ of the real Gold06 data indicating that it is an unlikely event. Figs. 2b and 3b show the corresponding plots obtained with the same dataset (Gold06) but the tested model is the PDL crossing $(w_0,w_1)=(-1.4,2)$ (best fit $\omm=0.34$) instead of \lcdm $(w_0,w_1)=(-1,0)$. In this case we find $z_c=1.01$, $N_c=15$ (Fig. 2b) and the probability of $z_{mc} \leq z_c$ ($N_{mc} \geq N_c$) is $32\%$ (Fig. 3b). Clearly, the Gold06 data are significantly more consistent with this cosmological model according to the BND statistic.

A similar analysis as the one shown in Figs. 2, 3 is shown in Figs. 4, 5  for the case of the Union08 dataset (307 datapoints). The results are similar and consistent with those based on the Gold06 dataset. For \lcdm we find $z_c = 0.53$ ($N_c = 137$) which is reproduced ($z_{mc}\leq z_c$) only by $5.3\%$ of the corresponding Monte Carlo datasets. In contrast, the crossing redshift $z_c=1.37$ ($N_c=2$) obtained with the $(w_0,w_1)=(-1.4,2)$ (best fit $\omm=0.30$) is consistent with all of the corresponding Monte Carlo datasets.

The tension between \lcdm and recent datasets in the context of the BND statistic has been verified by comparing high redshift data with corresponding Monte Carlo data generated in the context of \lcdm. At lower redshifts ($z<1$) where predictions of different dark energy models for the luminosity distance tend to converge, we do not anticipate this tension to persist. In order to verify this anticipation, we have repeated the analysis by starting the procedure of generating $Q(N)$ from a lower redshift rather than the highest redshift point. We have chosen $z_s =0.8$ as the staring point and we have applied BND crossing statistic to the Union08 datasets testing the two cosmological models. For both \lcdm and PDL models the crossing redshift of $z_c=0.791$ ($N_c=3$) is obtained that is consistent with all of the corresponding Monte Carlo datasets (Fig. 6). Thus, despite of the decrease of the data errors at lower redshifts, it is clear that the tension between \lcdm and recent data in the context of the BND statistic is no longer present.

\begin{figure*}[ht]
\centering
\begin{center}
$\begin{array}{@{\hspace{-0.10in}}c@{\hspace{0.0in}}c}
\multicolumn{1}{l}{\mbox{}} &
\multicolumn{1}{l}{\mbox{}} \\ [-0.20in]
\epsfxsize=3.3in
\epsffile{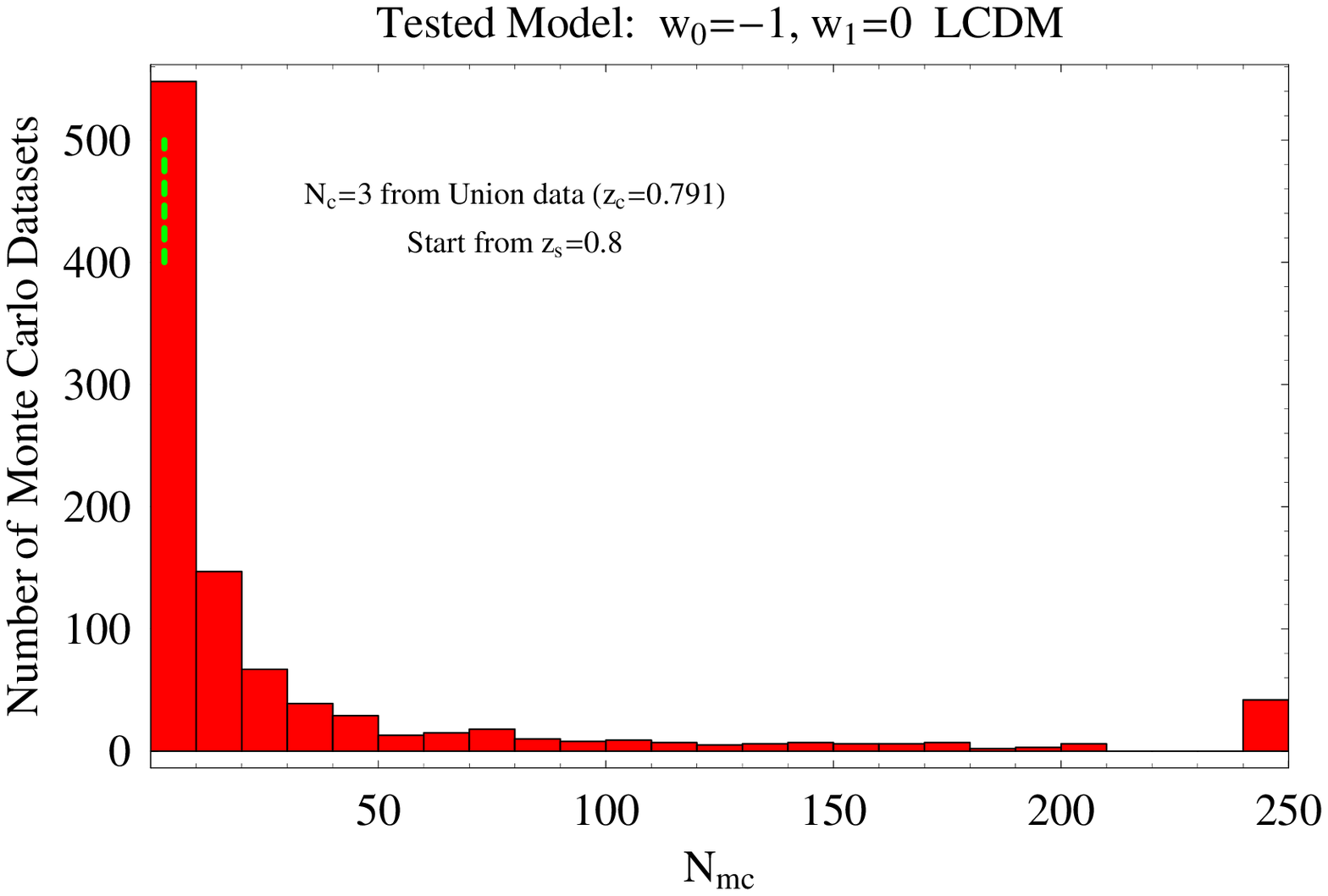} &
\epsfxsize=3.3in
\epsffile{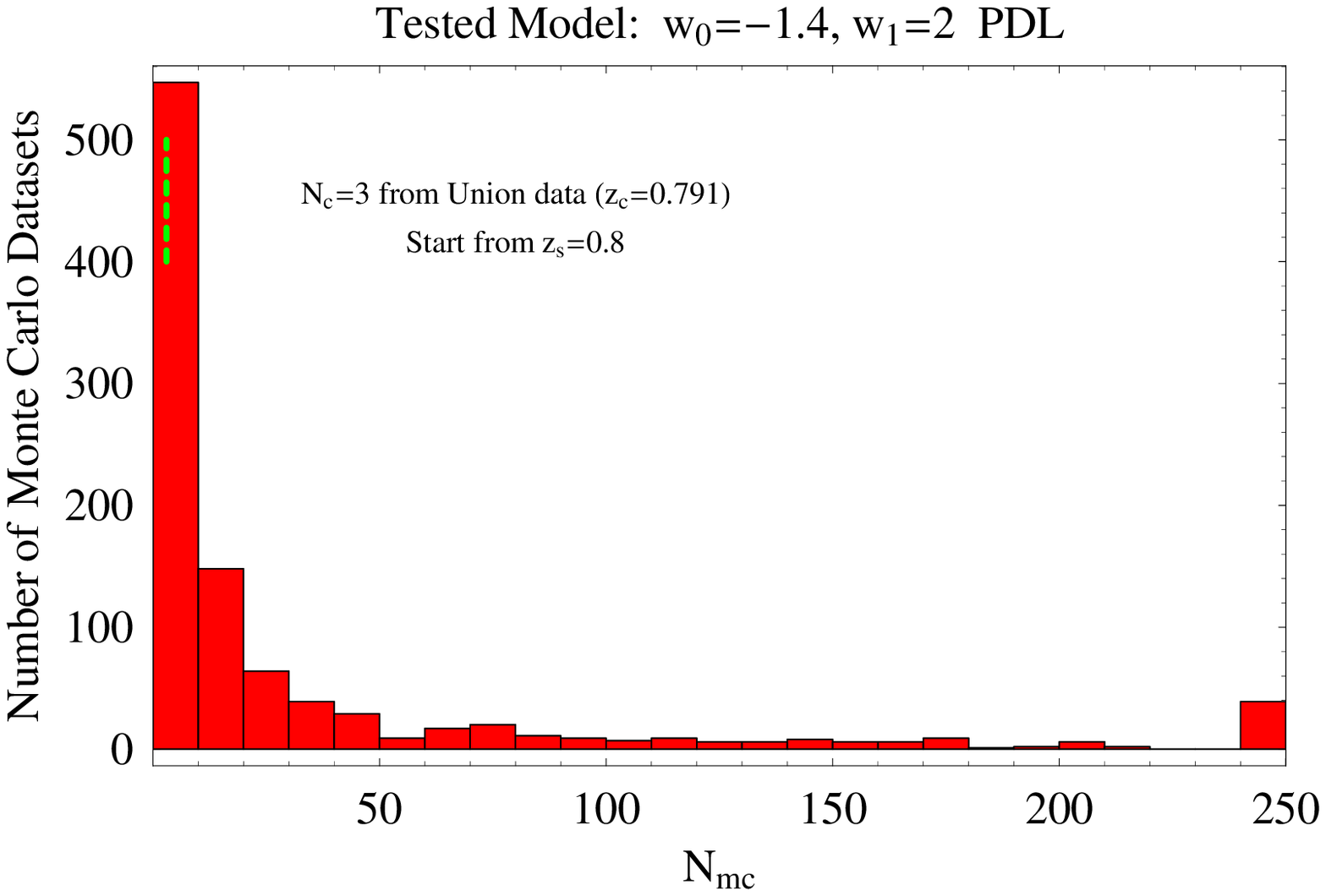} \\
\end{array}$
\end{center}
\vspace{0.0cm}
\caption{\small a: A histogram of the probability distribution of $N_{mc}$, using $z_s =0.8$ as the staring point instead of using the highest redshift point. Result obtained using Monte Carlo Union08 data under the assumption of a best fit \lcdm cosmological model ($\omm=0.29$). The thick green dashed line corresponds to the crossing redshift $z_c$ of the real Union08 data. Fig. b: Similar histogram for the PDL crossing model $(w_0,w_1)=(-1.4,2)$
(best fit $\omm=0.30$) instead of \lcdm. Notice that both models show a proper concordance to the data at these lower redshifts.}
\label{fig5a}
\end{figure*}

An interesting question to address is the following: `How robust is the derived tension between the recent SnIa data and \lcdm if there are additional systematic errors in the data due to a possible SnIa evolution?'. In order to address this question, we have added quadratically an additional error of $\sigma_{sys}=0.20$ to the error-bars of all data points in the Union08 sample (and also in its Monte Carlo realizations) and we have repeated the analysis. This amount of additional systematic error is comparable with the original error-bars of the data in Union08 sample. For \lcdm we found $z_c = 0.76$ ($N_c = 66$) which is reproduced ($z_{mc}\leq z_c$) by $12.1\%$ of the corresponding Monte Carlo data sets. It is clear that by assuming this additional systematic error, the consistency of \lcdm and Union08 data is increased with respect to $5.3\%$ consistency obtained from the actual data. In Fig. 7 we show the resulting histogram for this case. Clearly, the tension between data and \lcdm is reduced by including the additional errors but it is still beyond the $1\sigma$ level.


\begin{figure}[!t]
\hspace{0pt}\rotatebox{0}{\resizebox{.5\textwidth}{!}{\includegraphics{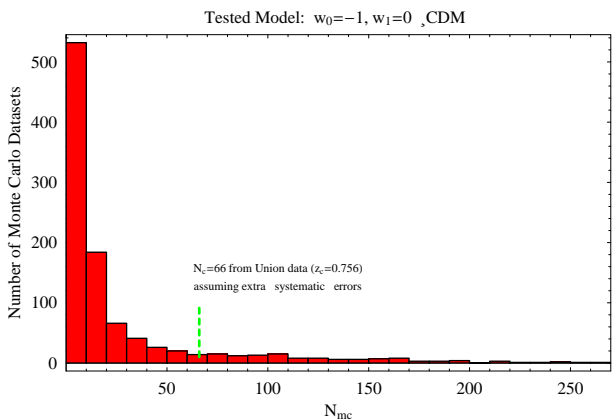}}}
\vspace{0pt}{\caption{\small A histogram of the probability distribution of $N_{mc}$ obtained using Monte Carlo Union08 data assuming additional systematic error of $\sigma_{sys}=0.2$ under the assumption of a best fit \lcdm cosmological model ($\omm=0.298$). The thick green dashed line corresponds to the crossing redshift $z_c$ of the real Union08 data with the assumed additional errors.}} \label{fig1a}
\end{figure}

\section{Conclusion}
We conclude that according to the BND statistic, the Gold06 and Union08 datasets have probability $2.2\%$ and $5.3\%$ respectively to have emerged in the context of the best fit \lcdm cosmology but the corresponding probabilities for the PDL crossing model $(w_0,w_1)=(-1.4,2)$ are larger than $30\%$. We have demonstrated that the identified tension between \lcdm and recent data is due to the data-points at high redshift that seem to be systematically brighter than the \lcdm predictions. At lower redshifts ($z\leq 0.8$) where the predictions of the various dark energy models for the luminosity distance tend to converge, we have verified that the revealed tension is no longer present (Fig. 6). Our result indicates a potential challenge for \lcdm cosmology and provides the motivation for obtaining additional SnIa data at high redshifts $z>1$. Such data, may confirm or disprove the anomalous high $z$ SnIa brightness which is mainly responsible for the low probability of the high $z$ SnIa data in the context of $\Lambda CDM$.

Clearly, the unexpected high $z$ brightness of SnIa can be interpreted either as a trend towards more deceleration at high $z$ than expected in the context of \lcdm or as a statistical fluctuation or finally as a systematic effect perhaps due to a mild SnIa evolution at high $z$. However, in view of the fact that a similar mild trend for more deceleration than expected at high $z$ is also observed in the context of standard rulers \cite{cross-evid,Percival:2007yw,data-cross}, the latter two interpretations are less likely than the first.

The data and the mathematica files used for the production of the figures may be downloaded from http://leandros.physics.uoi.gr/bnd.zip

\section*{Acknowledgements}
We thank R. Pain for useful discussions. This work was supported by the European Research and
Training Network MRTPN-CT-2006 035863-1 (UniverseNet).

\end{document}